\documentclass[12pt]{article}
\usepackage{amsfonts,latexsym}

\def\ignore#1{\ }

\newcommand{\m}[1]{ { $#1$} }

\newcommand{\beq}{ \begin{eqnarray}}
\newcommand{\eeq}{\end{eqnarray}}
\newcommand{\numeq}{\end{eqnarray}}

\newcommand{\half}{ {1\over 2} }

\newcommand{\beqs}{\begin{eqnarray}}
\newcommand{\eeqs}{\nonumber\end{eqnarray}}

\newcommand{\eps}{\epsilon}

\usepackage{graphics}

\refstepcounter{equation}

\begin{document}

\begin{titlepage}

\vspace{1in}
\begin{center}
{\LARGE
\textbf{A Model of Interacting Partons \\
\vspace{0.1in}
 for Hadronic Structure Functions \footnote{Received 
the American Physical Society's Apker Award for 1999.}}
}

\vspace{0.3in}

{\Large

by \\
\vspace{0.3in}
Govind S. Krishnaswami \\
Advisor: Professor S.G.Rajeev \\
Department of Physics and Astronomy, University of Rochester
Rochester, NY 14627, USA \\
\vspace{0.3in}
}

\vspace{1.8in}
\end{center}

{\large
\noindent Submitted as a senior thesis to the University of Rochester [in 
partial fulfillment of the requirements for the degree Bachelor of Science 
in Physics], May 1999}

\vspace{0.6in}

\vspace*{\stretch{1}}
\end{titlepage}

{\ }

{\ }

{\ }

{\ }

{\ }

{\ }

\begin{center}
{\large\bf Abstract}
\end{center}

{\ }

	We present a model for the structure of baryons in which the valence 
partons  interact  through  a linear potential. This  model can be
derived  from QCD in the approximation where transverse momenta are
ignored. We compare  the
valence quark distribution function predicted by our model with that
extracted  from  global fits to  Deep Inelastic
Scattering data. The only parameter we can adjust is the fraction of baryon
momentum  carried by valence partons.
Our predictions agree  well with data except for small values of the 
Bjorken scaling variable.

\pagebreak
\tableofcontents

\pagebreak

{\ }

{\ }

{\ }

\section{Introduction}
\label{introduction}

	This thesis is about the structure of the proton. The proton, along 
with the neutron is one of the constituents of the atomic nucleus. For a 
long time since their discovery, it was not known whether the proton and
neutron had substructure, and if so, what their constituents were like. The
Deep Inelastic Scattering Experiments \cite{friedman} of the early 1970s found that the
proton was made of point-like constituents called quarks or partons. 

	These experiments were similar in spirit to the alpha particle scattering
experiments of Rutherford, which established that the atom contains a
point-like nucleus. He scattered alpha particles against a thin gold foil
and found that there was a small probability for them to
scatter through wide angles. This would not be possible if the positive
charge in an atom was uniformly distributed. Moreover, the
scattering cross-section he measured was that of a point-like nucleus
carrying all the positive charge of the atom! Subsequent experiments showed
 that
the nucleus was made of protons and neutrons. Charge radii measured 
in elastic electron-proton scattering showed that the proton
was not elementary. What were its constituents like?

	The Deep Inelastic Scattering experiments scattered electrons
against protons. This time, the scattering was inelastic. The 
inclusive scattering
cross-section was measured and expressed in terms of `structure functions'.
The structure functions describe the structure of the proton. These
experiments used the electromagnetic force between the electron and the
constituents of the proton to study the `strong force' that held the proton
together. The electromagnetic force is mediated by the exchange of a photon.
By making the wave-length of the photon small enough, it was possible to
`look' deep within the proton.

	The startling discovery of the Deep Inelastic Scattering Experiments 
was that as long as one looked closely
enough, the structure functions did not depend on the wave-length of the
photon! The constituents of the proton did not have any length-scale
associated with them! This phenomenon is called scaling: the constituents of
the proton appeared point-like.

	The parton model was proposed by Bjorken, Feynman and others
\cite{parton} as a simple explanation of scaling in Deep Inelastic
Scattering. The proton was thought of as being made of point-like
constituents called partons. These were identified with the quarks which
were until then, hypothetical particles. Structure functions can be
expressed in terms of parton distribution functions. These describe the
distribution of partons within the proton.

	Soon after, Quantum ChromoDynamics (QCD), the fundamental theory of
strong interactions was discovered. Scaling was understood as a consequence
of asymptotic freedom in QCD \cite{af}. Small violations of scaling 
were predicted by
perturbative QCD. These have been confirmed by experiments. However, the
initial parton distributions cannot be calculated within perturbative QCD.
They have as yet not been understood theoretically. However, these parton
distributions are so important to high energy hadron collisions that they
have been measured and extracted from experimental data in great detail
\cite{cteq,mrst,grv}.

	Understanding the structure functions of quarks in a proton is not
unlike understanding the orbits of planets around the sun or the
wavefunctions of electrons in an atom. Understanding the latter has proven
extremely fruitful in all of science. In the case of the proton, we are
dealing with the strong force rather than the gravitational or
electromagnetic.

	In this thesis, we shall present a model of interacting partons to
explain the distribution of valence quarks in the proton. The quarks are 
assumed to be relativistic particles interacting with each other through a
linear potential in the `null' coordinate. Their momenta transverse to the
direction of the collision will be ignored in favour of the much larger
longitudinal momentum. That collinear QCD can describe hadronic structure
functions has also been proposed by others \cite{brodsky}.
The ground state wave function of the proton will be
the one that minimizes its energy. We will approximate the proton wave
function with a product of single parton wave functions. This is analogous to 
the Hartree approximation in Atomic Physics. These
approximations will allow us to determine the parton wave functions as the
solution to a non-linear integral equation. We shall study this equation in
several ways (both analytic and numerical) and obtain a fairly complete
quantitative picture of the valence quark distribution.

	Finally we will compare our predictions with the parton
distributions extracted from experimental data. Considering that we have
at-most one parameter to adjust (the fraction {\it{f}}, of proton 
momentum carried by the valence quarks) the agreement is quite good, except for
small values of the momentum fraction \m{x_{B}}. In this region, our
model is not expected to be accurate. We may not ignore sea-quarks, gluons
and transverse momenta.

{\ }

\centerline{\includegraphics{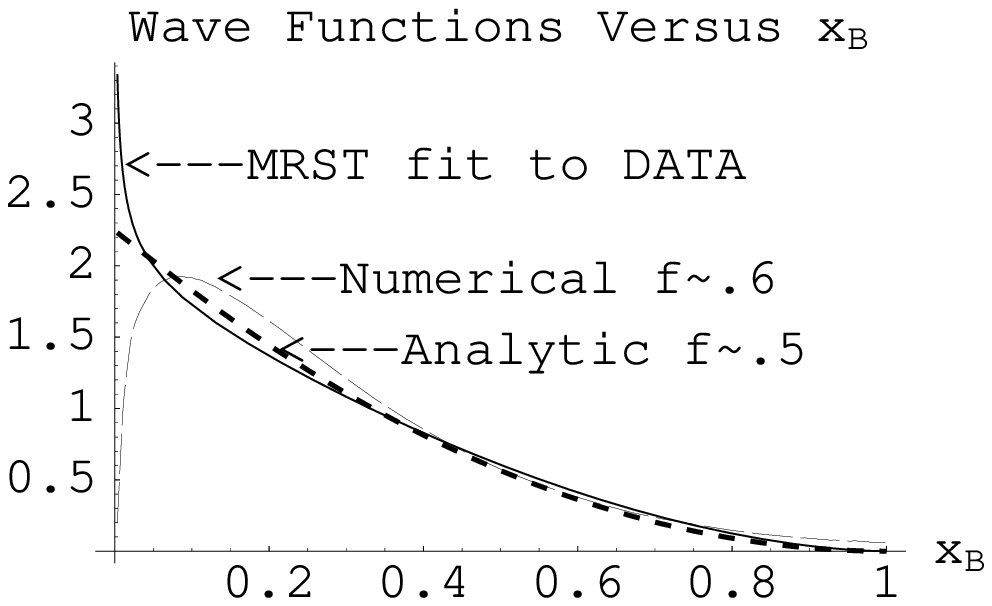}}

\noindent
{\small Figure 1. Comparison of the predicted valence parton wavefunction 
$\surd\phi(x_B)$ with the MRST \cite{mrst}
global fit to data. The wavefunction we predict goes to a non-zero 
constant at the origin. 
The `analytic' prediction is obtained as a variational 
approximation. The numerical solution is not reliable in this 
region of small $x_B$. The fit to data has a mild divergence at 
$x_B = 0$.

{\ }

{\ }

	These results have been described in a recent publication
\cite{govraj}. Furthermore, this interacting parton model can be derived
from QCD \cite{qhd} in the limit where transverse momenta are ignored.

	The question of deriving the spectrum and structure functions 
of hadrons from QCD
is an old and important one. The meson spectrum and wave functions of 
two dimensional QCD
were obtained by 't Hooft \cite{thooft} in 1974 by a clever summation 
of planar Feynman
diagrams in the large N limit. Perturbation theory works in the case of 
mesons since they are
described by small fluctuations from the vaccuum. However, the baryons 
remained elusive.
Later, Witten \cite{witten} suggested that the baryon can be described by a Hartree Fock approximation in the large N limit of QCD. He carried out this idea in a non-relativistic context. In the early 1980s, Skyrme's \cite{skyrme} idea that the baryon is a topological soliton was revived by Balachandran and others \cite{bal} and shown to be consistent with QCD. Rajeev \cite{qhd} developed Quantum HadronDynamics (QHD) in two dimensions. Two dimensional QHD is an equivalent reformulation of two dimensional QCD in terms of observable particles: the hadrons, rather than the quarks, which are confined within the hadrons. Baryons are the topological solitons of QHD while mesons are small fluctuations of the vaccuum. The large N limit of 2dQHD reproduces \cite{qhd} the meson spectrum and wave functions of 't Hooft. But it also allows us to
predict the structure of the baryon. Within a variational approximation, the structure of the baryon can be estimated by a non-linear integral equation \cite{qhd}. It was found that this integral equation also had a simple derivation in terms of the parton model \cite{govraj,qhd}. We present this parton model point of view here.

	This thesis is organised into 3 main chapters. Chapter~\ref{dis}
provides background, a discussion of Deep Inelastic Scattering and the
parton model. In Chapter~\ref{ivp} we present the interacting parton model and
derive the integral equation satisfied by the valence quark wave function.
Chapter~\ref{solution} discusses how we solve for the valence quark wave function and
comparison with experimental data. We do not assume much familiarity with particle physics.
We will often use terms that are specialized
to this branch of physics, but most of them are defined or explained at some
point.

\pagebreak

{\ }

{\ }

{\ }

\section{Deep Inelastic Scattering}
\label{dis}

	In electron-proton Deep Inelastic Scattering for instance, a high
energy electron scatters against a proton. The electron does not feel the
strong force and its weak interaction is dominated by the electromagnetic
interaction, which is mediated by the exchange of a high energy virtual photon. Thus
the kinematics is described in terms of two 4-vectors: $q^\mu$ the momentum of
the photon and $p^\mu$ the momentum of the proton. The photon momentum is space-like, 
$q^2 < 0$ while $p^\mu$ is time-like, $p^2 = M_{p}^2 > 0$. $M_{p}$ is the rest
mass of the proton. The scattering is inelastic. The proton disintegrates
producing several hadrons in the final state:
\begin{center}
	\m{ e P \to e X}
\end{center}
	If we sum over all possible final states X, we get the inclusive
deep inelastic cross-section, which is expressed in terms of two
dimensionless scalar `structure functions' $F_{1}$ and $F_{2}$. Being Lorentz
scalars, they can depend only on $p^2$, $p.q$ and $q^2$. $p^2 = M_{p}^2$ is 
fixed by the mass of the proton. For convenience, we may take our two independent
scalars as $Q^2 = - q^2$, and the dimensionless ratio $x_{B} =
\frac{Q^2}{2p.q}$, called the Bjorken scaling variable.

So \m{ F_{1,2} = F_{1,2}(x_{B},Q^2).}

	Q$^2$, being the square of the momentum transferred by the photon,
sets the energy scale of the experiment. Q is inversely proportional to the
wave length of the photon. An experiment at large Q$^2$ is therfore
looking deep inside the proton. If $Q^2 >> {1 \over a^2}$, we are
in the deep inelastic region. ${1 \over a} 
\sim 100 MeV$ is the inverse charge radius of the proton in its
rest frame.

	In the deep inelastic region, we will show that $x_B$ can be
thought of as the fractional momentum of the quark inside the proton that
scatters the photon. This will be made clear in the context of the parton
model to be discussed soon.

	The Deep Inelastic Scattering experiments of the early 1970s
\cite{friedman} showed that at sufficiently large $Q^2 >>
\Lambda_{QCD}^2$, the structure functions were approximately independent of
Q$^2$. They depended, not on the Lorentz Scalars $Q^2$ or $2p.q$ separately,
 but only on their ratio $x_B$. This phenomenon is called Bjorken scaling.

	The structure functions may be expressed in terms of parton 
distribution functions
\cite{cteqhandbook}. They directly describe the structure of the proton. The parton distribution functions $\phi_{a}(x_B,Q^2)$ are
momentum space probability distribution functions. For instance, the up
quark parton distribution function gives the probability density of finding an up quark carrying a fraction $x_B$ of proton momentum, when the proton is probed at the energy
scale Q$^2$. As with the structure functions, the parton distribution functions are essentially
independent of Q$^2$. Being probability distributions, they are the
absolute squares of the parton wave functions inside the proton.
The reduction from structure functions to parton distribution 
functions is made possible by ignoring certain correlations 
and the factorization 
theorem of perturbative QCD \cite{factorization,cteqhandbook}. 
Roughly speaking, the amplitude for the virtual photon to scatter off
the proton is expressed as a sum of products of two factors: the amplitude 
for it to scatter elastically off a quark of given momentum 
and the probability of finding a quark with that momentum in the
proton. While the elastic scattering off a quark of a given momentum
is calculable in perturbation theory, the probability of finding
such a quark in the proton is non-perturbative and yet to
be calculated from first principles.


\subsection{The Quark Model and Quantum ChromoDynamics}
\label{quarkmodel}

	The quark model, based on hadron spectroscopy suggested that for the
purpose of quantum numbers, the proton is a colorless combination of 2
flavours of quarks: up, up and down. The quarks are fermions. In addition to
quantum numbers like position, spin and charge, quarks carry flavour (up,
down, strange etc.) and color quantum numbers. Color is the analogue of
electric charge in the theory of strong interactions. The number of 
colors N is 3 in nature. Quantum ChromoDynamics (QCD) is the presently
accepted theory of the color force. The strong force is mediated by the exchange of spin 1 bosons called gluons. It was developed partly in analogy with
Quantum ElectroDynamics (QED) which is the theory of the electromagnetic
force. QCD is a quantum field theory with a non-abelian gauge symmetry.
While its basic framework is understood, it has proven very hard to solve
and make predictions about non-perturbative effects.

	The proton is just one of several particles which take part in
strong interactions. They are collectively called hadrons. Hadrons are
broadly classified according to their spin. Integer spin bosonic hadrons are
called mesons while the half integer spin fermions are called baryons. 
The proton and neutron
are the lightest and most common baryons. The pions are the lightest
mesons. Free quarks have never been observed in nature: they are confined
within the hadrons, which are colorless combinations of quarks.

	In the late 1960s and early 70s, the parton model for hadrons was
proposed by Bjorken, Feynman and others \cite{parton} to explain 
the phenomenon of scaling in Deep Inelastic Scattering. According to the
original parton model, in the deep inelastic region, the proton behaved as
though it was made up of essentially free point-like constituents called
partons. These partons were identified with the quarks (u, u and d in the
case of the proton).

	When probed at even higher energies (larger \m{Q^2})
and small \m{x_B,} it was found that there is a small probability of finding
anti-quarks (such as \m{\overline{u}} and \m{\overline{d}}) and even 
quarks of other flavours
(strange quarks for instance) inside the proton. These did not fit 
directly into the
original parton model. However, these additional probability distributions
were measured and are described in terms of a phenomenological extension of
the original parton model. We now speak of valence, sea, anti-quark and
gluon distributions in the proton. The valence quarks are the 3 quarks 
(up, up and down in the case of the proton) of the original parton model. 
They are named in a loose analogy with the valence electrons of the atom. 
One needs to probe deeper into the
proton (higher $Q^2$) before `sea' quarks and anti-quarks become significant.

	In this paper we shall primarily be interested in developing a model
that predicts the \m{x_B} dependence of valence quark wave functions. Since
this is the main goal, we shall ignore some other details and small
corrections, which though important and measurable, obscure the main point.
For instance, the wave function depends on the isospin of the baryon (proton
or neutron). We shall calculate a single valence quark wave function
ignoring isospin effects, which must therefore be compared to the isoscalar
combination of experimentally measured up and down quark wave functions in
the baryon. Correcting for isospin effects is not hard, but will not be
addressed here.

	In the physics of strong interactions, our experimental knowledge
and precision of measurements far exceeds our theoretical understanding in
most areas. In contrast with QED, where theory and experiment agree to an
unprecedented degree, in the non-perturbative regime of QCD we are barely at
the qualitative or 10$\%$ level!

\subsection{Deep Inelastic Scattering and the Parton Model}
\label{kinematics}

	Let us now return to the kinematics of deep inelastic collisions. Up
to now we described the scattering in Lorentz invariant language. To
understand the parton model of Feynman better, it will be useful to consider
the `infinite-momentum' frame. 

\centerline{\includegraphics{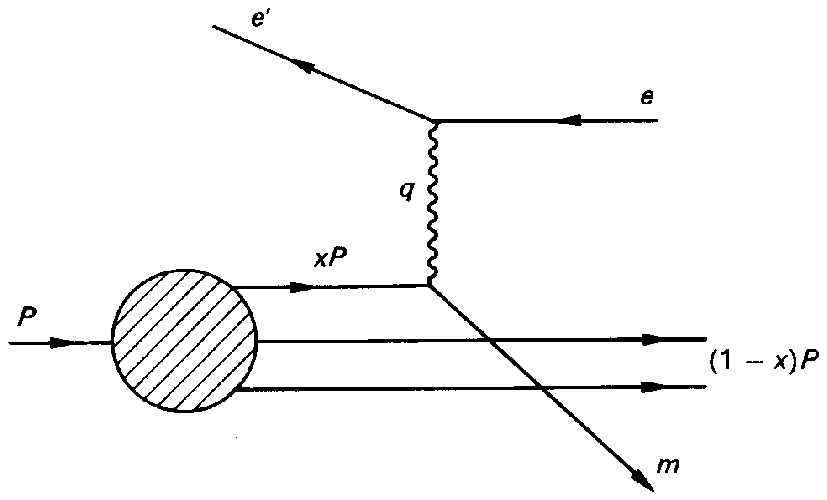}}

\noindent
{\small Figure 2. A parton model schematic of a Deep Inelastic Collision. 
Figure taken from \cite{perkins}.}

{\ }

	In this inertial reference frame, the proton
has a very large 3-momentum {\bf{P}} and we may ignore its rest mass in
comparison $p^\mu = (|\bf{P}|,\bf{P})$. The proton is thought of as
consisting of several point-like constituents, the partons. The 4-momentum
of a given parton is expressed as $xp^\mu$ where $0 \leq x \leq 1$ is
the fraction of proton momentum carried by the given parton. Deep Inelastic
Scattering is visualized as initially, the {\it{elastic}} scattering of the
photon against a parton. The partons are assumed to have negligible mass m.
This is followed by a complicated process of {\it hadronization} in which the 
partons recombine to form several hadrons in the final state X.

From our point of view, that of predicting parton distribution functions, 
there is an important simplification that Deep Inelastic Collisions allow.
The component of momentum in the direction of the
collision far exceeds those transverse to it. Therefore, it is reasonable to
ignore the transverse components of the parton momenta. It is this
approximation that makes the problem of predicting hadronic structure
functions tractable from a theoretical point of view. We are essentially
dealing with a problem in 2 space-time dimensions.

	The parton model also gives us a very useful physical interpretation
of the Bjorken scaling variable, $x_B$. $x_B$ was one of the two Lorentz
scalars used to parametrize the structure functions. We will show that
$x_B = \frac{Q^2}{2p.q}$ is the same as the fraction x of proton momentum
p carried by the parton that scatters off the photon. Momentum conservation
implies that $(xp + q)^2 = m^2$ where {\it{p}} is the 4-momentum of the 
proton, {\it{q}} the momentum of the photon and {\it m} the mass of the 
struck quark.
But m is negligibly small and $x^2p^2$ = $x^2M_p^2$ where $M_p$ is the mass of
the proton. In the deep inelastic region, $Q^2 >> M_p^2$ and therefore $x \sim
\frac{Q^2}{2p.q}$ which we recognize as the Bjorken scaling variable
$x_B$. Therefore we may interpret $x_B$ in the deep inelastic region
as the fraction of proton 4-momentum carried by a parton.

	This parton model interpretation of $x_B$ as the fraction of
proton momentum carried by a parton suggests that $0 \leq x_B \leq 1$
at least in the deep inelastic region. This is in fact true in general. It
follows from conservation of 4-momentum in a collision and the fact that the
proton is the lightest baryon. Since $x_B$ is a Lorentz scalar, it may be
evaluated in any inertial reference frame. In the lab frame, a photon of 
4-momentum 
$q^\mu~=~(\nu,\bf{q})$ collides inelastically with a proton of mass M$_p$ at 
rest. The invariant
(mass)$^2$ of the final state X is W$^2$.

\centerline{\includegraphics{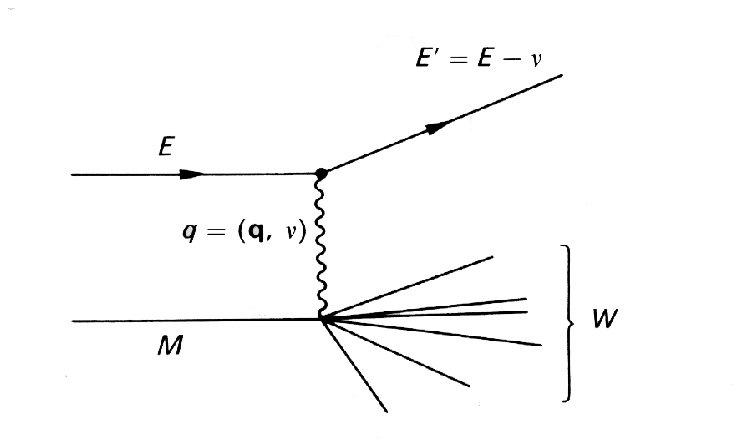}}
\noindent
{\small Figure 3. Kinematics of a Deep Inelastic Collision in the lab frame. Figure taken
from \cite{perkins}.}

{\ }

Even in an inelastic collision, 4-momentum is conserved:

\centerline{
\m{[(\nu,{\bf{q}}) + (M,{\bf{0}})]^2 = W^2}
}

\noindent Therefore, \m{q^2 + 2M\nu + M^2 = W^2,} where \m{q^2 = q^\mu q_\mu}. Writing
\m{Q^2} for \m{-q^2} we see that \m{Q^2 = 2M\nu - (W^2 - M^2)}. Now the proton
is the lightest baryon, while the final state X includes several heavier
hadrons, hence \m{W^2 \geq M^2} and \m{x_B = \frac{Q^2}{2M\nu} \leq 1}.
Moreover, since the energy transfer \m{\nu \geq 0} and \m{Q^2 \geq 0,} we have
\m{x_B \geq 0.} Therefore, \m{0 \leq x_B \leq 1} as desired.

While we are still in the lab frame, it will be good to point out
precisely what the deep inelastic limit is. We have $x_B =
\frac{Q^2}{2M\nu}$. The deep inelastic region of the $x_B - Q^2$ parameter
space is the limit of large $Q^2$ and $\nu$ keeping $x_B$ fixed.

	In the deep inelastic limit, the structure functions are essentially
independent of $Q^2$. They depend only on $x_B$. For instance, $F_2$ 
falls by about $50\%$ as $Q^2$ increases from 1 to
25 $GeV^2$, while the square of the elastic form factor falls by a
factor of $10^6$ over the same range! \cite{perkins}. This invariance of
the Structure Functions on the energy scale of the measurement, $Q^2$ is
known as Bjorken Scaling and was first observed in the Deep Inelastic
Scattering Experiments of the 1970s \cite{friedman}. The proton when 
probed at high energies (small
distances), appeared to consist of point-like constituents. 

\centerline{\includegraphics{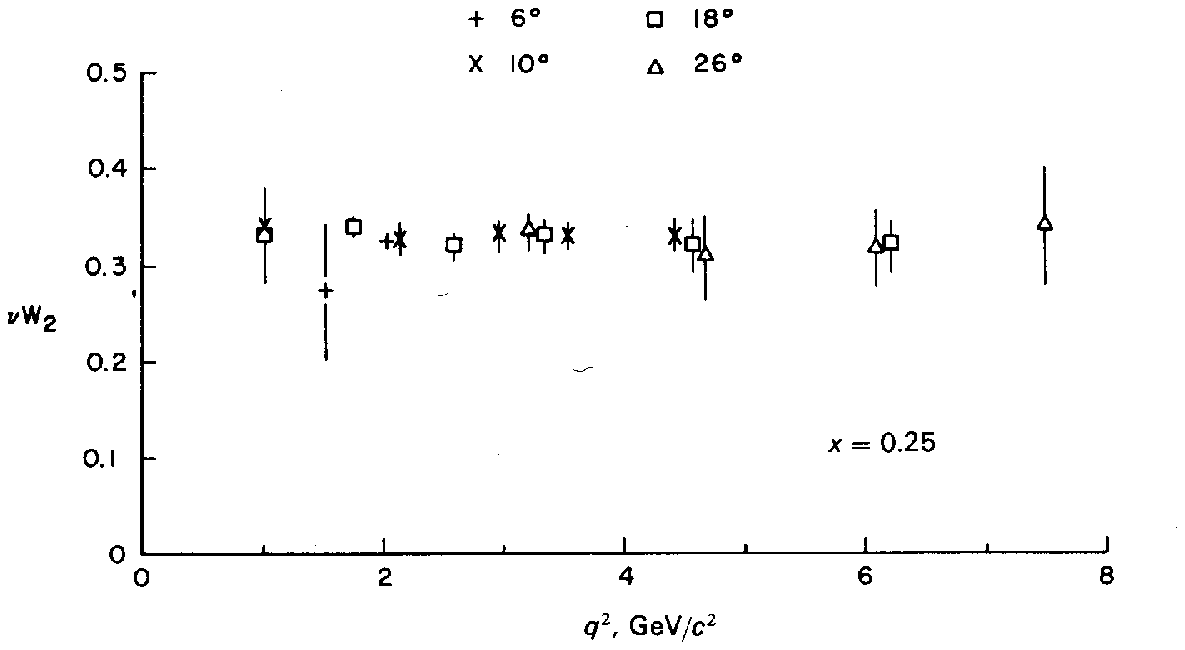}}

\noindent 
{\small Figure 4. $F_2$ as a function of Q$^2$ at $x_B$ = .25. For this choice 
of $x_B$, there
is practically no Q$^2$ dependence of the structure function, that is,
exact scaling. (After Friedman and Kendall \cite{friedman} (1972).)
Figure taken from \cite{perkins}.
}

{\ }

	Soon, however, it was found that scale invariance was only an
approximate symmetry: the structure functions had a very slow (logarithmic)
$Q^2$ dependence. The accurate prediction of the $Q^2$ dependence of the
structure functions is one of the major successes of perturbative QCD.
However, there is as yet no satisfactory theoretical understanding of the
$x_B$ dependence of the structure functions. It is essentially controlled
by non-perturbative effects. The fundamental theory of strong interactions
QCD has been in place for several years now, but has proven very hard to
solve even approximately. We shall describe a model which allows us to
calculate the $x_B$ dependence of hadronic structure functions. This model
can be derived as an approximation to the large N limit of two dimensional
QCD. N here is the number of colors. It was actually obtained as an
approximation to Quantum HadronDynamics(QHD), an equivalent reformulation 
of 2dQCD in terms of color invariant observables \cite{qhd}. However, we are
able to interpret and describe this model in terms of the parton model. It
is this parton model point of view that we shall describe in what follows.

	Before describing the interacting parton model, let us introduce 
a convenient coordinate system in two space-time dimensions: the null
coordinates.

\subsection{Null Coordinates}
\label{null}

	It will prove extremely convenient to use a null coordinate system
when describing the kinematics in the valence parton model. One useful
consequence of this is that the relativistic energy-momentum dispersion
relation \m{E = \pm \sqrt{p_1^2 + m^2}} will no longer involve complications
due to the square-roots. The sign of energy will be the same as that of
momentum! In QHD, the field variable M(p,q) depends on two space-time
points that are separeated by a null distance and thus causally
connected.

	As explained earlier, we will ignore the transverse momenta of the
partons. We may take the longitudinal momenta of the partons to be directed
along the $x^1$ axis. The two dimensional space-time position and momentum
in cartesian coordinates are $(x^0,x^1)$ and $(p_0,p_1)$. Rather than use
$p_1$ as the kinematic variable, we will use the null component of
momentum $p = p_0 - p_1$. `{\it{p}}' is the momentum along a null line in
2-dimensional Minkowski space. We will use $(p_0,p)$ as our coordinates.
This is not an orthogonal coordinate sytem. However, the simplification
achieved is that the mass-shell condition for a particle of mass m: 
$p_0^2 = p_1^2 + m^2$ is replaced by $p_0 = \half (p + \frac{m^2}{p})$
where $p = p_0 - p_1$ is the null momentum.

	Thus we see that for a particle with positive energy $p_0$, the
null momentum must be positive, unlike in cartesian coordinates.
Also note that the fraction x of proton 4-momentum carried by a parton 
is the same
as the fraction of corresponding null momenta. For brevity, we shall 
often refer to
null momentum just as momentum. There should be no confusion since we will no 
longer use the cartesian coordinate $p_1$.


\pagebreak

{\ }

{\ }

{\ }

\section{The Interacting Valence Parton Model}
\label{ivp}

\subsection{The Quark-Quark Potential}
\label{potential}

	We now describe a model for the structure of baryons in the language
of the parton model. In the original quark-parton model, the baryon is made
of N partons, which are essestially massless, free particles. N here is the
number of colors, which is 3. We shall keep N arbitrary for now. It is true
that when probed at very small distances (large momentum transfers, $Q^2$), the
quarks inside the proton appear to be free particles. This is called
asymptotic freedom \cite{af} in QCD. However, quarks are confined within 
the proton.
An isolated quark has never been produced. Therefore, even though the 
force that binds
quarks together is vanishingly small at small distances, it is non-zero at
intermediate distances and does not diminish at large distances. This is
in stark contrast to the electromagnetic and gravitational forces which
decrease with distance!

	If the partons in our model were indeed free, they would not bind to
form a proton. The only way the proton can have non-trivial structure
functions is for the quarks to interact. Since we are describing a many-body
system (N partons), it is simplest to assume an attractive two body potential
between the quarks.

	In Quantum Chromodynamics, the force between quarks is mediated by
the exchange of massless spin 1 bosons called gluons. Gluons are the
carriers of the color force just as photons are the quanta of the
electromagnetic force. When QCD is dimensionally reduced to 1+1 
dimensions \cite{qhd},
the gauge fields may be eliminated using their equations of motion. What
remains is effectively a long range quark-quark force which is given by a
linear two-body potential in the null coordinates.

	This linear potential can also be understood in other ways. Since
gluons are massless, their propogator in momentum space is $\frac{1}{q^2}$. 
This is the
same as the photon propogator. In 4 space-time dimensions, the photon
propogator corresponds to the $\frac{1}{|r-r^{'}|}$ Coulomb potential. 
However, since we are
in 2 space-time dimensions, we must use the appropriate Coulomb potential,
which is the Green's function of the Laplace operator (second derivative with
respect to x) in 1 space dimension. This is $\half |x-y|$. Notice that 
this potential is linear, attractive and confining.

	Moreover, there is evidence that the quark-quark potential is linear
as reported in \cite{rosner}. This interacting parton model can be 
derived as an
approximation to Quantum HadronDynamics (QHD) \cite{govraj,qhd}. In QHD, 
Lorentz
invariance leads to the choice of a linear potential. Translation 
invariance implies that
it can only depend on the absolute value of the separation \m{|x-y|}.

\subsection{Wave Function of the Proton}
\label{wavefn}

	We assume that the proton is made up of N
partons (quarks). Therefore, the proton is a relativistic quantum
many-body system. We will describe the proton in terms of a wave function
that will depend on the quantum numbers of the partons (momenta, spin,
color and flavour.)

	We are interested in knowing the baryon wave function in its ground
state, since that is where the proton spends most of its time. The strategy
will be to write down the energy of the baryon in the state $\psi$ and
minimize this energy with respect to different choices of $\psi$. The
configuration $\psi$ which minimizes the energy is its ground state wave
function. This is what we will compare with experimental data.

	The quarks are spin $\half$ fermions transforming under the fundamental
representation of the color group $SU(N)$. However, the proton itself is
colorless. (i.e. a color singlet) It 
must transform under the trivial representation of $SU(N)$. As mentioned
earlier, we will work in the null coordinates. Therefore the wave function of
a single parton is $\tilde\psi(a,\alpha,p)$. Here `a' labels the flavour
(up, down) and the spin quantum numbers of the parton. $\alpha$ labels color
and $p = p_0 - p_1$ is the null momentum of the parton. We use
$\tilde\psi$ to denote the momentum space wave function. The corresponding
position space wave function $\psi(a,\alpha,x)$ is the inverse Fourier
transform of $\tilde\psi(a,\alpha,p)$. Since the null momentum is always
positive, $\tilde\psi$ must vanish for negative p. We shall often refer to
this as `$\tilde\psi$ has only positive support'.

	We can think of $\tilde\psi$ as a joint probability density. It is
a probability density that depends on two discrete random variables `a' and
`$\alpha$' and one continuous random variable p, for any given parton. The 
proton is made up of N partons and has a wave function

\centerline{
\m{\tilde\psi(a_{1},\alpha_{1},p_{1}; \cdots ; a_{N},\alpha_{N},p_{N})}
}
 
Here
$a_{i}, \alpha_{i}, p_{i}$ are the spin, flavour, color and null momentum
of the $i^{th}$ parton. Since the quarks are fermions, the proton wave function
should be completely anti-symmetric under interchange of a pair of quarks,
by the Pauli exclusion principle. However, since the baryon is invariant
under color, the wave function must be completely anti-symmetric in color
alone.
\begin{center}
\m{
\tilde\psi(a_{1},\alpha_{1},p_{1}; \cdots ;a_{N},\alpha_{N},p_{N}) = 
\frac{\eps_{\alpha_{1}, \cdots , \alpha_{N}}}{\sqrt{N!}} \tilde\psi(a_{1},
p_{1};\cdots ; a_{N},p_{N})
}
\end{center}

	This means the wave function is completely symmetric in spin,
flavour and momentum quantum numbers. Therefore, once color has been
factored out, the partons behave like a collection of bosons. 

\subsection{Hamiltonian of the N-Parton System}
\label{hamiltonian}

	What is the energy of the proton in the state
$\tilde\psi(a_{1},p_{1}; \cdots ; a_{N},p_{N})$? We will express
the energy as a sum of kinetic and potential energies. Consider the case $N =
1$,i.e. a proton consisting of a single parton with wave function
$\tilde\psi(a,p)$. The `effective mass' of the parton is $\mu_{a}$,
which could possibly depend on what its flavour is. We shall return to a
discussion of the effective mass later. The relativistic kinetic
energy-momentum relation expressed in terms of null-momentum is 

\begin{center}
\m{p_{0} = \half (p + \frac{\mu_{a}^2}{p}).}
\end{center}

The probability density that the parton indeed has null momentum p and
spin-flavour `a' is \m{|\tilde\psi(a,p)|^2}. Therefore, the expectation value
of the kinetic energy of this single parton baryon is 

\begin{center}
\m{
\sum_{a} \int_0^{\infty} {\half (p + \frac{\mu_{a}^2}{p})
|\tilde\psi(a,p)|^2 \frac{dp}{2\pi}}
}
\end{center}

\noindent It follows that the kinetic energy of a baryon with N partons is 

\begin{center}
\m{
\sum_{a_{1},a_{2}, \cdots,a_{N}} \int_0^{\infty} \sum_{i=1}^{N} \half (p_i
+ \frac{\mu_{a_i}^2}{p_i}) |\tilde\psi(a_{1},p_{1}; \cdots ; a_{N},p_{N})|^2
\frac{dp_{1} \cdots dp_{N}}{(2\pi)^N}
}
\end{center}

\noindent (We shall choose our normalization constants for Fourier transforms so that a
factor of $\frac{1}{2\pi}$ consistently appears in the measure of every
momentum space integral.)

	The potential energy is most easily expressed in position space. As
explained earlier, the partons interact through a two-body
potential. i.e. the potential energy of a pair of quarks at null
coordinates x and y is $g^2 v(x-y)$, where g is a coupling constant with
the dimensions of mass. The potential \m{v} is linear in the separation
between the quarks $v(x-y) = \half |x-y|$. This yields a potential energy

\begin{center}
\m{\frac{\tilde{g}^2}{2} \sum_{a_1,...,a_N} \int_{-\infty}^{\infty} \sum_{i \ne j} v(x_i - x_j) |\psi(a_1,x_1;\cdots;a_N,x_N)|^2 dx_1 \cdots dx_N}
\end{center}

\noindent for the system of N partons. The factor of $\half$ is to avoid double
counting. $\psi(a_{1},x_{1}; \cdots ;a_{N},x_{N})$ is
the position space wave function of the baryon. It is the inverse Fourier
transform of the momentum space wave function.

\begin{center}
\m{
	\psi(a_1,x_1;\cdots a_1,x_N)=\int_0^\infty 
\tilde\psi(a_1,p_1;\cdots a_N, p_N)
 e^{i\sum_j p_jx_j}{dp_1\cdots dp_N\over (2\pi)^N} 
}
\end{center}

	We said that $\mu_{a}$ is the effective mass of `a' parton of flavour
a inside the proton. This must be explained. A quark is confined within the
proton and cannot be isolated as a bare particle. Therefore, its bare mass
is not a directly measurable quantity. All that matters is its effective mass
$\mu_{a}$ when confined within the proton. However, one can make an indirect
measurement of of the `bare' quark masses. At high energies, the strong
force is essentially zero and the electro-weak interactions of these
essentially free quarks depends on their bare masses. These have been
measured and are called `current' quark masses. The current quark mass of
the up and down quarks, which are the valence quarks in the proton, are 
about 5 and
8 MeV/c$^2$. This must be contrasted with the energy scale of the strong
interactions $\Lambda_{QCD} \sim$ 100 MeV. We see that the bare up and
down quark masses are negligible in comparison. In QCD, the limit of zero
quark mass is the limit of Chiral Symmetry. It is probably best to think of the quark mass parameters $m_a$ as measuring the extent to which Chiral Symmetry is broken in the theory. In nature, the quark masses are not exactly zero, but they are zero to a good first approximation. It is this limit that we shall mostly be interested in while comparing our results with experimental data.

	Returning to the question of effective and bare quark masses, we
mention that a similar situation exists in the case of an electron in a
metal. It has an effective mass that is different from the mass of a free
electron. The term in the energy that involves the effective mass may be
re-expressed as the sum of a term involving the bare mass alone and a term
involving the coupling constant. This latter term is often referred to as
the self-energy term. In QED, which is an interacting theory of the electron,
the `bare' electron can emit and re-absorb photons. This part of the interaction
can be thought of as renormalizing the bare mass of the electron, giving 
rise to a self-energy

	However, these analogies are only to set the context. They cannot be
pushed too far, in the present case,
the square of the effective mass is infact negative in the limit of chiral
symmetry! In this case, the self energy term really arises in 2 dimensional QCD
when a quartic operator in the potential energy is normal ordered \cite{qhd}. The
resulting expression for the effective mass in terms of the bare mass is $\mu_a^2 = m_{a}^2 - \frac{\tilde{g}^2}{\pi}$, where $\tilde{g}^2 = Ng^2$. This expression has been derived in the literature \cite{thooft,hagen}.

	Adding the potential and kinetic energies, the total energy of the
proton is

\beq
	{\cal E}_N(\tilde\psi)&=&\sum_{a_1\cdots a_N}\int_0^\infty \sum_{i=1}^{N}\half[p_i+{\mu_{a_i}^2\over
p_i}]|\tilde\psi(a_1,p_1;\cdots a_N,p_N)|^2{dp_1\cdots
dp_N\over (2\pi)^N}\cr
 & &+ \half g^2 \sum_{a_1\cdots a_N}\int_{-\infty}^\infty
\sum_{i\neq j}
v(x_i-x_j)|\psi(a_1,x_1;\cdots a_N,x_N)|^2 dx_1\cdots
dx_N.\nonumber
\eeq

The wave function that minimizes this energy subject to the constraint that
the total probability is 1 is the ground state baryon wave function. Here
 the norm of $\tilde\psi$ is given by

\begin{center}
\m{
||\tilde\psi||^2 = \sum_{a_1,\cdots,a_N} \int_0^{\infty} |\tilde\psi(a_1,p_1;\cdots ;a_N,p_N)|^2 \frac{dp_1 \cdots dp_N}{(2\pi)^N}
}
\end{center}

\subsection{The Hartree Ansatz}
\label{hartree}

	The minimization of the above energy can be thought of as a problem
in many-body theory. We don't have a general solution to such a problem and
must look for approximation methods. An important simplification can be made
since we are describing the ground state of a bosonic system, since we have
factored out the totally anti-symmetric color part of the wave function.
Mean field theory gives a good approximation to the ground state where each
boson moves in the mean field created by the others. Moreover, in the ground
state, each of the bosons can be assumed to occupy the same single particle
state. Therefore, in the mean field approximation, we make the ansatz:

\begin{center}
\m{
\tilde\psi(a_1,p_1;\cdots;a_N,p_N)=
\delta(\sum_ip_i-P)\prod_{i=1}^N\tilde\psi(a_i,p_i).
}
\end{center}

In the language of probability distributions, we are assuming that the
joint probability distribution of the N partons is well approximated by the
product of N independent (identical) single parton distributions. Since the
null momenta are positive we impose the condition that the null momenta of
the partons add up to that of the baryon ({\it P}). Therefore, 
the single parton
wave functions must vanish for $p < 0$ and $p > P$. 
(Infact, the condition is weaker, all we need is the delta function factor
on the momentum sum above. But the difference is a correlation which is
suppressed in the limit of a large number of partons, N.)
Thus we are ignoring quark-quark correlations. When comparing with data,
we will express the wave function in terms of the dimensionless fractional 
momentum $x = \frac{p}{P}$. In Section~\ref{kinematics} we showed that this fractional momentum is nothing but the Bjorken scaling variable $x_B$ in the deep inelastic limit.

	In practice, this is a very reasonable assumption. It underlies most
of the phenomenological description of Deep Inelastic Scattering. The very
fact that one can speak of an up quark momentum distribution in the proton,
without mentioning the simultaneous down quark momentum at which it was
measured means that it is a good approximation to ignore correlations.

	Such a mean field approximation is common in many-body physics and
is probably best known in the context of the Hartree approximation of
atomic physics. There, we are solving the Schr\"{o}dinger equation for a many
electron atom. The number of electrons in a neutral atom is the same as the atomic number Z. The mean field approximation can be thought of as a large-Z approximation \cite{qhd}. The fact that it works well even for Helium when Z~=~2 encourages us to use it here for N~=~3. For a more precise analogy, see the
appendix to \cite{qhd}.

\subsection{Momentum Sum Rule and Boundary Conditions}
\label{sumrule}

	There is one further constraint that the single particle wave
function must satisfy: the momentum sum rule. The expectation value of the
total momentum of all the valence partons must equal the fraction $f$ of the
baryon momentum actually available to them

\begin{center}
\m{
        N\int_0^P p|\tilde\psi(p)|^2{dp\over 2\pi}= fP.
}
\end{center}

\noindent The valence quarks do not carry all the momentum of the baryon, 
as in the
original parton model. The parton distributions extracted from Deep
Inelastic Scattering data show that roughly half the momentum of the baryon
is carried by the valence quarks. The rest is in the gluons, sea quarks and
anti-quarks. We shall see that the fraction \m{f} is the only parameter 
on which the parton distributions predicted by our model depend.

	The energy per parton is therefore:

\beqs
	E&=&\sum_a\int_0^P \half[p+{\mu_a^2\over
p}]|\tilde\psi(a,p)|^2{dp\over 2\pi}+\cr
  & & \half{\tilde{g}^2}\int_{-\infty}^\infty v(x-y)
\sum_a|\psi(a,x)|^2\sum_\beta|\psi(\beta,y)|^2dxdy.
\eeqs

\noindent with the potential $v(x-y) = \half |x-y|$ as explained in Section~\ref{potential}. The
single parton wave function is determined by minimizing the energy per
parton subject to the constraints: $||\tilde\psi|| = 1$ and
$\tilde\psi(a,p) = 0$ unless $0 \leq p \leq P$. We will impose the boundary
condition that the wave function vanish as $p \to P$. This is because the 
probability density of a single parton carrying all momentum of the baryon 
must be zero.
The behavior of the wave
function as $p \to 0$ will be worked out in detail in the next chapter 
(Section~\ref{frobenius}). It will turn out that for a positive quark 
mass $m > 0$, the wave function vanishes at the origin too.
In the limiting case $m = 0$, the wave function goes to a non-zero value 
at the origin.

	The condition that the momentum space wave function vanish outside
[0,P] looks rather different in position space. It says that
$\psi(a,x)$ must be the value along the real line, of an entire
function. This constraint can be hard to implement especially if we look for
numerical solutions. Hence we work in momentum space.

	We also make a comment on the support of the wave function. We have 
imposed the constraint that $\tilde\psi(p)$ vanish for
$p > P$. i.e. the null momentum of a parton must
not exceed that of the baryon. However, the momentum sum rule states that
$N \int_0^P {p|\tilde\psi(p)|^2 \frac{dp}{2\pi}} = fP$, where $f$ is the
fraction of baryon momentum carried by the N valence partons. But this is
just N times the mean (first moment) of the single parton distribution! We
see that the single parton wave function is primarily concentrated around
the region $p = \frac{fP}{N}$. For large enough N, it should be
reasonable to allow the wave functions to be non-zero beyond P, since they
are vanishingly small for large p anyway. Thus we may replace the upper
limit of integration P, by $\infty$ especially while obtaining analytic
approximations. Note that we are not saying that P $\to \infty$. 
This is not a Lorentz invariant statement. P is the null momentum 
of the baryon and has a fixed value in any given inertial reference frame.


\subsection{Positivity of the Energy}
\label{positivity}

	We may reformulate this constrained minimization problem as the
solution of a certain non-linear integral equation. Before proceeding with
this, let us make some slight simplifications. We will set all the parton
masses equal to each other. In nature, the up and down current quark masses
are both roughly zero. Further, let us look for a single parton wave
function that is non-zero for only a single spin-flavour index `a' :
$\tilde\psi(a,p) = \delta_{a,1}\tilde\psi(p)$.

	This breaks the U(M) invariance of our model spontaneously. This
symmetry can be restored later by the collective variable method as in the
theory of solitons, though we shall not address these issues here.

	With these simplifications, the energy is given by:
\beqs
	E&=&\int_0^P \half[p+{\mu^2\over
p}]|\tilde\psi(p)|^2{dp\over 2\pi} + 
   \half{\tilde{g}^2}\int_{-\infty}^\infty v(x-y)
|\psi(x)|^2|\psi(y)|^2dxdy.
\eeqs

We take a moment to check that the energy is positive (and hence
bounded below), so that its minimization is a well posed problem! This is
immediate since the integrands of both the kinetic and potential energies
are positive functions for $\mu^2 = m^2 - \frac{\tilde{g}^2}{\pi} 
\geq 0$. The physically interesting region is the limit of zero quark mass $m
\to 0$. This corresponds to a negative value of $\mu^2$. This case is 
more subtle, the
kinetic $(\int_0^P \half[p+{m^2\over p}]|\tilde\psi(p)|^2{dp\over 2\pi})$ and potential energies $(\half{\tilde{g}^2}\int_{-\infty}^\infty v(x-y)|\psi(x)|^2|\psi(y)|^2dxdy)$ remain positive while the self energy term $(\int_0^P \half[{-\tilde{g}^2\over \pi p}]|\tilde\psi(p)|^2{dp\over 2\pi})$ is
negative. However, it can be shown that this self energy term is cancelled
by a portion of the potential energy, when expressed in momentum space. In 
practice, this will not be a problem
since we will approach the physically interesting region from positive
values of $\mu^2$ and see that a sensible limit exists.


\subsection{The Energy in Momentum Space}
\label{momspace}

	The constraint on the position-space wave function is that it be the boundary value of an entire function. The same constraint is much simpler in momentum space: it must vanish outside the interval [0,P]. Therefore, as mentioned in Section~\ref{sumrule}, it will be convenient to express the potential energy in momentum space. The energy per parton is

\beqs
	E&=&\int_0^P \half[p+{\mu^2\over
p}]|\tilde\psi(p)|^2{dp\over 2\pi} + 
   \half{\tilde{g}^2}\int_{-\infty}^\infty v(x-y)
|\psi(x)|^2|\psi(y)|^2dxdy.
\eeqs

The potential energy is best understood in the language of electrostatics.
It is just the electro-static potential energy of a charge density 
$\rho(x) = \tilde{g}|\psi(x)|^2$ in one space dimension, where the Green's
function is $\half |x-y|$. $\tilde{g}$ here is analogous to the unit of 
electric
charge. The Poisson integral formula then gives the electrostatic potential

\begin{center}
\m{ V(x) = g \int_{-\infty}^{\infty} |\psi(y)|^2 \frac{|x-y|}{2} dy
}
\end{center}

Rather than refer to V(x) as the electrostatic potential, we must call it
the mean potential due to the partons. It is not to be confused with the
2-body potential $v(x-y)$ between the quarks! V(x) then satisfies Poisson's
equation  $V''(x) = \tilde{g} |\psi(x)|^2$. The boundary conditions are $V(0)
= \tilde{g} \int_{-\infty}^{\infty} {|\psi(y)|^2 \frac{|y|}{2} dy}$ and 
as $|x| \to
\infty, V(x) \to \frac{\tilde{g}|x|}{2}$, which is just the asymptotic 
form of the
potential of a charge $\tilde{g}$ localized around the origin.

	We shall rewrite all of this in momentum space, but will have to be
careful since we will find that the mean potential $\tilde{V}(p)$ has a
$\frac{-1}{p^2}$ singularity at the origin in momentum space. The momentum
space integrals we get will therefore be singular and we will define them as
Finite Part integrals in the sense of Hadamard \cite{hack}. These
`Finite Part' prescriptions are not to be thought of as a witch-craft that
allows one to assign finite values to divergent integrals, but as a way of
restating the above boundary conditions in momentum space. Ultimately, it is
the physics that determines what the correct definition of an apparently
divergent quantity is. As we shall see, these definitions will turn out to
be very natural and motivated by general principles such as analytic
continuation. It is therefore likely that these are the `correct'
definitions in other contexts too.

	We first rewrite Poisson's equation for the mean potential in
momentum space by Fourier transforming: $-p^2 \tilde{V}(p) = \tilde{W}(p)$ where $\tilde{W}(p)$ is the convolution $\int_0^P {\tilde\psi(p+q) \tilde\psi(q) \frac{dq}{2\pi}}.$ We see that the mean potential is singular at the origin: $\tilde{V}(p) \to \frac{-1}{p^2}$ as p $\to 0$. It will be useful to note that
$\tilde{W}(0) = 1, \tilde{V}(-p) = \tilde{V}^{*}(p)$ and $\tilde{W}'(0) =
-\half |\tilde\psi(0)|^2$. Also, by a choice of phase, the wave function $\tilde\psi(p)$ may be taken to be real. Therefore $\tilde{V}(p)$ is real and even.

Therefore, the energy in momentum space is:

\begin{center}
\m{ E[\tilde\psi] = \int_0^P {\half (p + \frac{\mu^2}{p}) |\tilde\psi(p)|^2
\frac{dp}{2\pi}} + \frac{\tilde{g}}{2} {\cal FP} \int {\frac{dp
dq}{(2\pi)^2} \tilde{V}(p) \tilde\psi(q) \tilde\psi^{*}(q-p)}
}
\end{center}

\begin{center}
where \m{\tilde{V}(p) = \frac{-\tilde{g}}{p^2} \int_0^P {\tilde\psi^{*}(p+r)
\tilde\psi(r) \frac{dr}{2\pi} }
}
\end{center}

\noindent We see that the integrand in the potential energy has a
\m{\frac{1}{p^2}} singularity and we must define it as a `Finite Part'
({\cal FP}) integral. We will postpone a discussion of Finite Part 
integrals till we
actually have to evaluate them in the next chapter.


\subsection{Integral Equation for the Ground State Wave Function}
\label{integeqn}

	We must minimize the above energy functional with respect to the
constraint that the norm of $\psi$ be 1. This is equivalent to minimizing
$F[\tilde\psi] = E[\tilde\psi] + \lambda ||\tilde\psi||^2$. $\lambda$
here is the Lagrange multiplier enforcing the constraint. Since the
functional we are minimizing is quartic in $\psi$, $\lambda$ is {\it{not}} the
same as the ground state energy. Therefore the groundstate wave function
satisfies the equation $\frac{\delta F[\tilde\psi]}{\delta\tilde\psi} = 0$, 
which when written out more explicitly reads

\begin{center}
\m{
\half (p + \frac{\mu^2}{p}) \tilde\psi(p) + \tilde{g} FP \int_0^P
{\tilde{V}(p-q) \tilde\psi(q) \frac{dq}{2\pi} } = \lambda\tilde\psi(p)
}
\end{center}

\noindent Since $\tilde{V}$ is quadratic in $\tilde\psi$, we see that this is a
non-linear singular integral equation for $\tilde\psi$.

	In the next chapter, we will explain how the singular integrals are
defined and obtain a solution to the equation. We must mention that there is
a well-developed theory of linear integral equations, even those with
Cauchy-type singularies \cite{hack}. But the non-linearities and severity
of the singularities in our case mean that we will have to develop special
techniques to deal with our integral equation.

	The strategy will be to understand the ground state wave function
from several different points of view. No single technique may give us the
exact solution. We will develop approximation methods which are valid in
distinct special cases. By patching these together, we will obtain both a
qualitative and quantitative understanding of the solution.

\subsection{Parameters in the Model}
\label{paraneters}

	Let us conclude this chapter with a discussion of the parameters in 
the theory. $\tilde{g}$ is the coupling
constant of the theory. It has dimensions of mass and is related to the
coupling constant of two-dimensional QCD by $\tilde{g} = g^2 N$. 
$g$ is related to the
coupling constant of the four dimensional theory through a multiplicative 
factor of the order of the inverse transverse size of the proton. This
arises when we `integrate over' transverse coordinates to
arrive at this `effective' two-dimensional model. Though we do not 
predict the value of the coupling constant, we will see that
it cancels out when we calculate the valence quark distribution.

N is the number
of colors, which is the same as the number of valence quarks. Most of the
approximations we have made become exact when $N \to \infty$. We will use
the value N = 3 when comparing with data.

	$\mu$ is the `effective' quark mass and is related to the current
quark mass by $\mu^2 = m^2 - \frac{g^2}{\pi}$. We will be interested in the 
limit of chiral symmetry when $m \to 0$. The energy expressed above
has logarithmic divergences in both the potential and self energies when m
is exactly zero. However, we shall define the theory as the limit m $\to 0$. A
Frobenius analysis will show that the valence quark wave function has a
sensible limit as m $\to 0$. This can be thought of as an infrared regulation.
But we emphasize that there is no real divergence in the problem,
requiring renormalization.
In any event, nature is balanced on a knife-edge at a small, positive value
of m!

	The limit where m is small keeping $\tilde{g}$ fixed is the
extreme relativistic limit. The binding energy of the system far exceeds the
rest-mass of the partons. We will also have occasion to consider the
non-relativistic limit in which $\mu >> \tilde{g}.$ This is not the
physically interesting case as it corresponds to a baryon made up of heavy
quarks. However, in this limit the integral equation will reduce to a
non-linear ordinary differential equation  and will give us another 
way of studying the solutions.

	$\lambda$ is the Lagrange multiplier enforcing the constraint on
the norm of $\tilde\psi.$ It has no direct physical meaning and is not
equal to the energy of the state $\tilde\psi.$

	P is the null momentum of the baryon, which in its restframe, 
is equal to its
rest-mass. $\tilde\psi$ is non zero only for $0 \leq p \leq P.$ But 
when we express the wave function in terms of the dimensionless Lorentz
invariant variable $x_{B} = \frac{p}{P}$, we will see that the structure
functions do not depend on which value of P we used (i.e. which inertial
reference frame we picked). $\tilde{g}$ too will cancel out. One can think
of P as being measured in units of $\tilde{g}$. When the dimensionless
structure functions are expressed in terms of the dimensionless variable
$x_{B}$, they no longer depend on $\tilde{g}$.

	The only free parameter in the theory then, is the fraction $f$ of
baryon momentum carried by the valence partons. It enters through the
momentum sum rule $N \int_0^P {p |\tilde\psi(p)|^2 \frac{dp}{2\pi}}~=~fP$.


\pagebreak

{\ }

{\ }

{\ }

\section{Determination of the Valence Quark Wave Function}
\label{solution}

\subsection{Estimation of the Ground State using Variational Ansatzes}
\label{variational}

	The variational principle of quantum mechanics allows us to estimate
the ground state energy and wave function of a quantum mechanical system.
The ground state is the one that actually minimizes the energy. Therefore,
by minimizing the energy within a sub-class of functions in the Hilbert
space, we may obtain an upper bound for the ground state energy and also a
qualitative understanding of the ground state wave function. In practice, a
judicious choice of a 1 or 2 parameter family of variational ansatzes
motivated by physical principles can often come spectacularly close to the
true ground state energy.

	The energy per parton is given by

\beqs
	E&=&\int_0^P \half[p+{\mu^2\over
p}]|\tilde\psi(p)|^2{dp\over 2\pi} + 
   \half{\tilde{g}^2}\int_{-\infty}^\infty v(x-y)
|\psi(x)|^2|\psi(y)|^2dxdy.
\eeqs

	As a first approximation, we may work on the interval $[0,\infty)$
with a function that decays as $p \to \infty$, as justified in 
Section~\ref{sumrule}. A
simple choice is $\tilde\psi_a(p) = Cpe^{-ap}$ with $a > 0$, 
$p \geq 0.$ Here `a' is a variational parameter controlling 
the rate of decay of the wave function in 
momentum space. C is fixed by normalization. In position space, this positive 
momentum function has an analytic continuation to the upper half plane 
and a double pole at $x = -ia$: $\psi(x) = \frac{C^{'}}{(x+ia)^2}$.

	We may calculate the energy of such a state and minimize with
respect to a. It is clear on dimensional grounds that the term $\int_0^{\infty} \half p |\tilde\psi(p)|^2 \frac{dp}{2\pi}$ scales like $\frac{1}{a}$ while all other terms in the energy scale like a. (`a' has dimensions of inverse momentum or length.) Thus there is a value of `a' at which the energy is minimized among this class of functions.

\begin{center}
\m{
E_{ground~state} \leq \half \sqrt{I_1(m^2 I_2 + \tilde{g}^2I_3)},
}
\end{center}

\noindent where $I_{1}$, $I_{2}$ and $I_{3}$ are positive dimensionless pure numbers
obtained from the integrations.

	We have also tried another class of variational ansatzes: $\tilde\psi(p) = C p^{\alpha} (1-p)^{\beta}.$ This time, we work on the finite interval $[0,P]$ and set $P = 1$ for
convenience. Thus p is the fraction of baryon momentum carried by the quark. This is the same as the Bjorken Scaling variable $x_{B}$. The energy calculations in this case are more formidable 
but can be performed in terms of hypergeometric functions. Moreover, we work
in momentum space and use the definitions of singular integrals, which will be
presented in Section~\ref{finitepart}. The results, however, are encouraging and easy
to state. Among the choices $\alpha, \beta = 1,2,3$ we find that the energy is
minimized for the choice $\alpha = 1$ and $\beta = 2$, even as $m \to 0$. 
There is a further decrease in energy when we consider wave functions 
of the form $\tilde\psi(p) = C p^{\nu} (1-p)^2$ for fractional values of $\alpha$, $0 < \alpha < 1$. Our conclusion is that in the limit of chiral symmetry (zero current quark mass), the energy is minimized in the limit where $\alpha \to 0$. We thus have a qualitative understanding of the momentum space wave function as $m \to 0$. It goes to a non-zero constant as $p~\to~0$ and then falls off roughly as $(1-p)^2$ as $p \to 1$.

	We shall see that this picture is re-inforced and made more 
precise by the other approximation methods we develop to study the 
integral equation for $\tilde{\psi}$.

\subsection{Non-relativistic Limit}
\label{nonrel}

	The non-relativistic limit is the limit where the binding energy of
the system is small compared to the rest mass of the quarks. It can be
achieved by making the quark mass large compared to the coupling constant 
$(m >> \tilde{g} )$. In the case of the proton, this is not a very interesting
limit since the up and down quark masses are essentially zero. However, we
will be able to understand our system from a slightly different perspective
this way. Besides, it also gave us a way of checking our numerical
procedures. The expression for the energy is:

\begin{center}
\m{
E = \int_0^P \half (p + \frac{\mu^2}{p})|\tilde\psi(p)|^2 + \frac{\tilde{g}^2}{2} \int_{-\infty}^{\infty} |\psi(x)|^2 \frac{|x-y|}{2} |\psi(y)|^2 dx dy.
}
\end{center}

	The relativistic energy-null momentum dispersion relation is 
$p_0 = \half (p + \frac{\mu^2}{p})$, $p > 0$. We see that $p_{0}$ has a minimum
when $p = \mu$. In the extreme
case of no interaction at all $(\tilde{g} = 0)$, the total energy is just the
kinetic energy. In this case the energy is minimized when all the
null-momentum of the parton is concentrated at the value $p = \mu$. In other
words, $|\tilde\psi(p)|^2$ is a delta function at $p = \mu$. 

	The effect of having a small, but non-zero interaction $\tilde{g} > 0$
is to broaden out the ground state wave function. In any case, it is
concentrated around the minimum of the dispersion curve. Therefore we may
replace the dispersion relation by its quadratic approximation in the
neighbourhood of its minimum, $p = \mu$. The result should not be surprising.
We get the new dispersion relation 

\begin{center}
$p_{0} = \mu + k^2/2\mu$,
\end{center}

\noindent where $k = p - \mu$ is the shifted momentum variable and 
$\mu = \sqrt{m^2 - \frac{g^2}{\pi}} \sim m$ is the mass of the heavy quark.

	We immediately recognize this as the non-relativistic energy of a
particle of mass $\mu$ and momentum k. Putting this into the expression for
the energy per parton, we have
\begin{center}
\m{
E = \mu + \int \frac{k^2}{2\mu} |\tilde\chi(k)|^2 \frac{dk}{2\pi} + \frac{\tilde{g}}{2} \int_{-\infty}^{\infty} V(x) |\chi(x)|^2 dx.
}
\end{center}

where $\tilde\chi(k) = \tilde\psi(\mu+k)$ and $\chi(x)$ is its Fourier
transform. V(x) is the effective potential satisfying Poisson's equation with boundary conditions as in Section~\ref{momspace}:

\begin{center}
\m{
V''(x) = g |\chi(x)|^2
}
\end{center}

	To minimize this energy we vary with respect to $\chi(x)$ and 
impose the constraint $||\chi|| = 1$. The result is a Schr\"{o}dinger-like
equation.

\begin{center}
\m{
\frac{-1}{2\mu} \chi''(x) + \frac{\tilde{g}}{2} V(x) \chi(x) = \lambda \chi(x)
}
\end{center}

\noindent It is non-linear, since V itself depends on $\chi$. However, 
it can be
solved by iteration. We start with a guess for the ground state wave
function $\chi_{0}(x)$ and calculate $V_{0}(x)$ by solving Poisson's equation.
Using this approximate effective potential $V_{0}(x)$, we solve the 
eigenvalue problem

\begin{center}
\m{
\frac{-1}{2\mu} \chi_1''(x) + \frac{\tilde{g}}{2} V_0(x) \chi_1(x) = \lambda \chi_1(x)
}
\end{center}

\noindent for the first iterate wave function $\chi_{1}(x)$. 
$\chi_{1}(x)$ is used to
determine a new effective potential $V_{1}(x) = \tilde{g} \int_{-\infty}^{\infty} |\chi_1(x)|^2 \frac{|x-y|}{2}dx$. (This is the solution of Poisson's equation satisfying the boundary conditions given in Section~\ref{momspace}). The process is iterated till it converges
(or until successive iterates are the same up to numerical errors, if one is working numerically). The resulting
ground state $\chi(x)$ may be transformed back to momentum space and
re-expressed in terms of $p = k + \mu$. We thus obtain an approximate ground
state wave function and energy for the N parton system in the
non-relativistic limit. We shall not discuss the details here.

	We mention the procedure primarily because it gives us a 
general method of obtaining approximate solutions to certain 
non-linear equations. It is essentially this procedure of iteration that
will be used in the numerical solution to our non-linear integral equation
for the ground state parton wave function. It will ofcourse be more
complicated since we have a non-linear integral equation with singularities.


\subsection{Finite Part Integrals}
\label{finitepart}

	We have mentioned the need to define singular integrals where the
integrand has a $\frac{1}{q^2}$ singularity. For instance, consider the 
integral
equation for the ground state wave function $\tilde\psi(p)$:

\begin{center}
\m{
\bigg[\half(p+{\mu^2\over p})-\lambda\bigg]\tilde\psi(p)+{\tilde
g^2}{\cal FP}\int_0^P\tilde{V}(p-q)\tilde\psi(q){dq\over 2\pi}=0,
}
\end{center}

\noindent where $\tilde{V}(p)=-{1\over p^2}\int_0^P
\tilde\psi^*(p+q)\tilde\psi(q){dq\over 2\pi.}$

	Hence the integrand in the equation for $\tilde\psi(p)$ is singular.
We will define it as a `Finite Part' integral in the sense of Hadamard
\cite{hack}. In general, all our singular integarls can be reduced to
those of the form

\begin{center}
\m{
{\cal FP} \int_0^P \frac{f(q)}{q^2} dq
}
\end{center}

\noindent where $f$ itself is not singular at the origin. In fact, we will assume that $f$
is $C^{1}$ at the origin from the right. This means that $f$ possesses a
continuous right-hand derivative at the origin. We shall rewrite the above
integral as 

\begin{center}
\m{
{\cal FP} \int_0^P \frac{f(q)}{q^2} dq = \int_0^P \frac{f(q)-f(0)-qf'(0)}{q^2} dq + f(0) {\cal FP} \int_0^P \frac{dq}{q^2} + f'(0) {\cal FP} \int_0^P \frac{dq}{q}
}
\end{center}

	The first integral on the right is non-singular at the origin and is
an ordinary Riemann Integral, since we have subtracted out the singular
parts. However, the last two integrals on the right still need to be
defined. They are of the form ${\cal FP}\int_0^P \frac{q^\nu}{q^2} dq$.
We know that for $\nu > 1$, this is just a Riemann Integral and has
the value $\frac{P^{\nu-1}}{\nu-1}$ We shall now define it even for 
$\nu < 1$ by analytically continuing this formula.

\begin{center}
i.e. \m{{\cal FP}\int_0^P \frac{q^\nu}{q^2} dq = \frac{P^{\nu-1}}{\nu-1}} for \m{\nu \ne 1.}
\end{center}

	This clearly runs into trouble for $\nu = 1$. In this special case, we
shall define it to be the average of the values for $\nu = 1 + \eps$ and $\nu =
1 - \eps$ as $\eps \to 0$.

	This is easily seen to imply that ${\cal FP}\int_0^P \frac{dq}{q} = Log(p)$. Therefore 
we have the following definition:

\beqs
{\cal FP}\int_0^P \frac{q^\nu}{q^2} dq =  \frac{P^{\nu-1}}{\nu-1} 
\hspace{.5in}\nu \ne 1.
\cr
{\cal FP}\int_0^P \frac{q^\nu}{q^2} dq =  Log(P) \hspace{.5in} \nu = 1.
\eeqs

	We must mention a few cautionary remarks. The definition for $\nu = 1$
violates the change of variable formula for the scale transformation $q \to \lambda q$ by the amount $Log\lambda$. We must take this into account
when making a scale transformation in the case $\nu = 1$.
	For $\nu \leq 1$, the Finite Part integral of a positive function
may be negative. For example

\begin{center}
\m{{\cal FP}\int_0^1 \frac{dq}{q^2} = -1.} and \m{{\cal FP} \int_{-\infty}^{\infty} \frac{dq}{q^2}  = 0.}
\end{center}

	However, we also note that these definitions reduce to the familiar
Riemann integrals whenever they exist in the usual sense. Furthermore, the
usual Cauchy Principal value for integrals with a simple pole that does not
lie at a limit of integration continues to hold. In particular then,

\begin{center}
\m{{\cal FP} \int_{-\infty}^{\infty} \frac{dq}{q} = 0}
\end{center}

	We emphasize that these definitions are necessary since we would
like to rewrite position space potential energy integrals in terms of
momentum variables. These complications arise because the integral kernel of
the Green's function $\frac{|x|}{2}$ expressed in momentum space takes the form
$\frac{-1}{p^2}$. These definitions ensure that the same integrals evaluated 
in position and momentum space give the same results. For instance, the
position space mean potential $V(x)$ satisfies Poisson's equation
$V''(x) = |\psi(x)|^2$ along with a pair of boundary conditions.
Poisson's equation in momentum space for the mean potential $\tilde{V}(p)$ is 

\begin{center}
\m{
-p^2 \tilde{V}(p) = \int_0^P \tilde\psi^{*}(p+r) \tilde\psi(r) \frac{dr}{2\pi}
}
\end{center}

\noindent The boundary conditions translate to rules for integrating the 
$\frac{1}{p^2}$ singularity appearing in $\tilde{V}(p)$.

	As an illustration and check, let us show that our definitions do
indeed imply the equality of the Green's functions expressed in position
and momentum space. We would like to prove that

\begin{center}
\m{
-\frac{|x|}{2} = {\cal FP} \int_{-\infty}^{\infty} \frac{1}{p^2} e^{ipx} \frac{dp}{2\pi}
}
\end{center}

	The proof will involve using our definition for Finite Part Integrals
since the integrand on the right is singular. We will use contour
integration to complete the proof.

	For x = 0, both sides are zero according to our definition. 
It suffices 
to prove the stated result for $x > 0$ since both sides are even functions 
of x as is easily checked.

\begin{center}
\m{
{\cal FP} \int_{-\infty}^{\infty} \frac{1}{p^2} e^{ipx} \frac{dp}{2\pi} = 
\int_{-\infty}^{\infty} \frac{e^{ipx}-1-ipx}{p^2} \frac{dp}{2\pi} + {\cal FP} \int_{-\infty}^{\infty} \frac{dp}{2\pi p^2} + ix {\cal FP} \int_{-\infty}^{\infty} \frac{dp}{2\pi p}
}
\end{center}
\hspace{1.9in}
\m{ = I_{1} + I_{2} + I_{3.}}

	$I_{3}$ is just the Cauchy principal value and is zero as observed
before. $I_{2}$ is also zero by the Finite Part prescription. $I_1$ is 
the Riemann integral of an entire
function of p along the real axis. It may be evaluated using a semicircular
contour in the upper-half plane. After making some estimates, we have
$I_{1} = \frac{-x}{2}$ for $x > 0$. Therefore, the desired result follows.


\subsection{The Small p Behavior of the Valence Quark Wave Function}
\label{frobenius}

	We shall describe a Frobenius-type analysis that gives us the
behavior of the wave function $\tilde\psi(p)$ for small positive p. The
integral equation for the wave function is:

\begin{center}
\m{
\bigg[\half(p+{\mu^2\over p})-\lambda\bigg]\tilde\psi(p)+{\tilde
g^2}{\cal FP}\int_0^P\tilde{V}(p-q)\tilde\psi(q){dq\over 2\pi}=0
}
{\ }

where \m{
	\tilde{V}(p)=-{1\over p^2}\int_0^P
\tilde\psi^*(p+q)\tilde\psi(q){dq\over 2\pi.}
}
\end{center}

Keeping only the singular terms for small p, we have:

\begin{center}
\m{
\half (m^2 - \frac{\tilde{g}^2}{\pi}) \frac{1}{p} \tilde\psi(p) + \tilde{g} {\cal FP}
\int_{p-P}^{p} \tilde{V}(q) \tilde\psi(p-q) \frac{dq}{2\pi} = 0
}
\end{center}

$\tilde{V}(q) \sim \frac{-1}{q^2}$ for small q. 
Now we assume a power law behavior $\tilde\psi(p) \sim p^{\nu}$ 
for small $p > 0$ and derive an equation 
for $\nu$.

\begin{center}
\m{
\half (\frac{m^2}{\tilde{g}^2} - \frac{1}{\pi}) = {\cal FP} \int_{1-\frac{p}{P}}^1 \frac{(1-y)^\nu}{y^2} \frac{dy}{2\pi}
}
\end{center}

where $y = \frac{q}{p}$. Since $p << P$, we have

\begin{center}
\m{
\frac{\pi m^2}{\tilde{g}^2} -1 = {\cal FP} \int_0^1 \frac{(1-y)^\nu + (1+y)^\nu}{y^2} dy
+ \int_1^{\infty} \frac{(1+y)^\nu}{y^2} dy
}
\end{center}

The first of these integrals is singular and we evaluate it according to the
definition given earlier. The result is a transcendental equation for
$\nu$:

\begin{center}
\m{
\frac{\pi m^2}{\tilde{g}^2} -1 = \int_0^1 \frac{(1+y)^\nu + (1-y)^\nu -2}{y^2} dy -2 + \int_1^{\infty} \frac{(1+y)^\nu}{y^2} dy
}
\end{center}

\noindent which is of the form \m{\pi m^2/\tilde{g}^2 - 1 = h(\nu).} It is 
easily seen that for m = 0, $\nu = 0$ is a solution. In the limit of
zero current quark mass, the critical wave function tends to a constant at
the origin. Calculating $h(\nu)$ shows that for a positive quark mass m,
the wave function goes to zero at p = 0 and rises like a power law
$\tilde\psi(p) \sim p^{\nu}$, $\nu > 0$. The following plot shows $h(\nu)$.
The solution $\nu$ for a given value of $m \geq 0$ is the point at which the
horizontal line $\pi \frac{m^2}{\tilde{g}^2} - 1$ intersects the curve.

{\ }

\centerline{\includegraphics{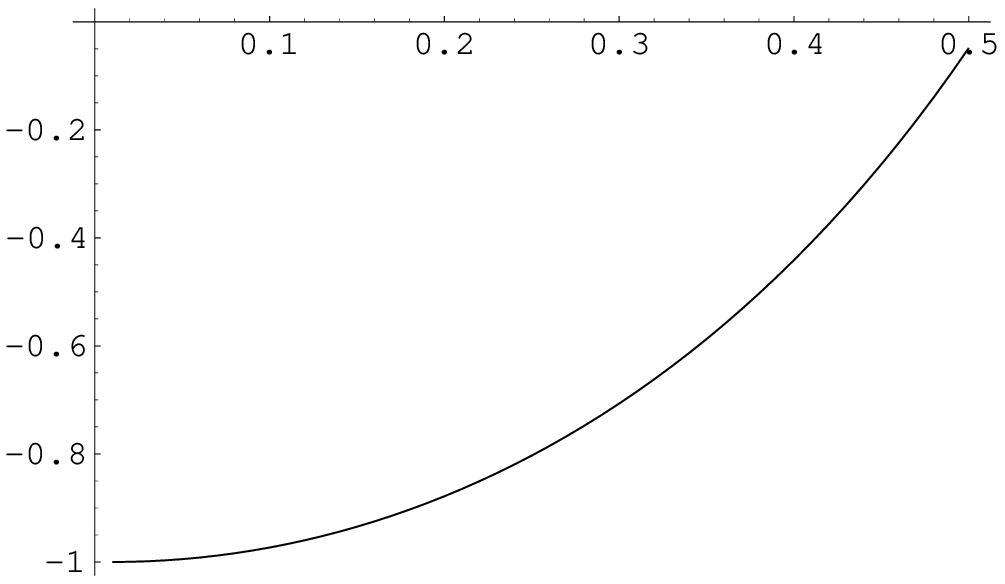}}

\noindent 
{\small Figure 5. h($\nu$) as a function of $\nu$. The solution $\nu$ 
for a given
 value of $m \geq 0$ is the point at which the
horizontal line $\pi \frac{m^2}{\tilde{g}^2} - 1$ intersects the curve.
Notice that for m = 0, $\nu$ = 0 is a solution: in the limit of
Chiral Symmetry, the wave function goes to a non-zero constant at the origin.}

{\ }

	Our conclusion that the critical momentum space wave function 
tends to a non-zero
constant as $p \to 0$ implies that it is discontinuous at the origin.
Therefore, the critical position space wave function decays like 
$\frac{1}{x}$ at infinity.
This is very reasonable from another point of view. In the soliton model
\cite{skyrme} the baryon is thought of as being made up of an infinite
number of pions. It has been shown that this picture is consistent
with QCD  \cite{bal}. But in that case, it should be possible to
approximate the baryon wave function with the meson wave function for large
spatial x. At large distances we are essentially far away from the proton. 
The meson
wave function has been calculated by 't Hooft \cite{thooft}. In the 
limit $m\to 0$
the 't Hooft meson wave function too has a discontinuity in momentum
space and consequently decays like $\frac{1}{x}$ as $x \to \infty$, just as we
obtain here.

	This Frobenius-like analysis is very useful. It gives us a clear 
quantitative understanding of the small p behavior of the wave function in
the physically interesting region $m \to 0$. We have that as $m \to 0$,
$\tilde\psi(p) \sim p^{\nu}$ with $\nu \to 0$. i.e. the wave function rises
sharply near the origin. At the critical point $m = 0$, the wave function goes
to a non-zero constant at $p = 0$. We had already got a scent of this behavior
from our variational estimates! Together with the numerical solution, which
will be valid except for small p, we will have almost a complete
understanding of the valence quark wave function.


\subsection{Numerical Solution}
\label{numericalsoln}

	The numerical solution of this
problem  is {\it{not}}  straightforward since the kernel of the integral
equation is singular. We need a reliable method of 
numerical quadrature for  integrals such as 

\begin{center}
\m{
	{\cal FP}\int_0^P f(p,q)\rho(q)dq
}
\end{center}

when the weight function \m{\rho(q)} has a singularity at \m{q=0} like
\m{1\over q^2.} 
We
need to subdivide the interval
\m{[0,P]=\cup_{r=1}^{n}[b_r,b_{r+1}]} into subintervals. Within
each subinterval we choose a set of points \m{q_{jr}, j=1,\cdots
\nu_r.} We approximate the integral by a sum 

\begin{center}
\m{
	{\cal FP} \int_0^P f(p,q)\rho(q)dq=\sum_{jr}w_{jr}f(q_{jr}).
}
\end{center}

The weights \m{w_{jr}} are determined by the condition that within
each subinterval \m{[b_r,b_{r+1}]}, the integral of a polynomial of
order \m{\nu_r-1} is reproduced exactly:

\begin{center}
\m{
 {\cal FP}\int_{b_r}^{b_{r+1}} q^k
\rho(q)dq=\sum_{j=1}^{\nu_r}w_{jr}q^k_{jr.}
}
\end{center}

This is the usual method of numerical quadrature \cite{hack,chandra} 
except that the
integral on the left hand side is singular for \m{b_r=0} and
\m{k=0,1.} In these cases we can evaluate the left hand side analytically as
the finite part in the sense of  Hadamard. 
(The main difference from
the usual situation is that  the moments
on the left hand side of the above equation are not all positive.) 
The weights are then
determined by solving the above system of linear equations.

Given an approximate mean potential \m{\tilde{V_s}(p)} we can convert
the linear integral equation 

\begin{center}
\m{
	\half[p+{\mu^2\over p}-2\lambda_s]\tilde\psi_{s+1}(p)+{\tilde
g^2}{\cal FP}\int_0^\infty\tilde{V_s}(p-q)\tilde\psi_{s+1}(q){dq\over 2\pi}=0
}
\end{center}

into a matrix eigenvalue problem by the above method of
quadrature. We use the ground state eigenfunction so determined to
calculate numerically the next approximation \m{\tilde{V}_{s+1}} for  the mean
potential. This process is  iterated until the solution converges.
Having determined the wavefunction, we must impose the momentum sum
rule to determine \m{\tilde{g}.} We used Mathematica to implement this
numerical procedure. An approximate analytic solution 
 was used as a   starting point for the iteration.


\subsection{Comparison with Experimental Data}
\label{experiment}

Now we turn to the question of the comparison of our model with data from
Deep Inelastic Scattering.  It is
customary to describe the parton distributions as functions of the
Bjorken scaling variable $0\leq x_B\leq 1$ which is the fraction of 
the null component of momentum
carried by each parton. This means we must rescale momenta to the
dimensionless variable $x_B = \frac{p}{P}$. The probability density of a parton
carrying a fraction $x_B$ of the momentum is then 
\begin{center}
\m{
	\phi(x_B)= {P\over 2\pi}|\tilde\psi(x_B P)|^2.
}
\end{center}
(The factor of $1\over 2\pi$ is needed because $\phi(x_B)$ is
traditionally normalized to one with the measure $dx_B$ rather than
$dx_B\over 2\pi$.)

It is important to note that the only dimensional parameter in our theory,
$\tilde g$, cancels out of the formula for $\phi(x_B)$: it only
serves to set the scale of momentum and when  the wavefunction is
expressed in terms of the  dimensionless variable
$x_B$, it cancels out. The wave function only depends on the
ratio ${m^2 \over {\tilde g}^2}$ and is independent of $\tilde g$
in the limit of zero current quark mass. 
We have set $\mu^2$ to the critical value (within numerical errors),
which is the value corresponding to chiral symmetry; i.e., zero current
quark mass. The number of colors we fix at $N=3$.
Thus the only free parameter in our theory is  the fraction $f$ of the baryon
momentum carried by the valence partons. The parameters $N$ and
$f$ appear in the combination $N_{\rm eff}={N\over f}$.

We have ignored 
the  ispospin  of the quarks in the above discussion. 
We should therefore compare our structure functions  with the isoscalar
combination of the valence quark distributions of a
baryon, $\phi(x_B)$. 
It is not difficult to take into account the effects of isospin.

The parton distributions have been extracted from scattering data by several groups of physicists \cite{cteq,mrst,grv}. In the figure below we plot our wavefunctions and compare them to that extracted from data by the MRST collaboration, at $Q^2=1$ GeV$^2$. We agree remarkably well with data except for small values of $x_B$. The agreement is best when the fraction of the baryon momentum carried by the valence partons is about $f~\sim~0.5$. 

{\ }

\centerline{\includegraphics{labelmrstvar.eps}}

\noindent
{\small Figure 1. Comparison of the predicted valence parton wavefunction 
$\surd\phi(x_B)$ with the MRST \cite{mrst}
global fit to data. The wavefunction we predict goes to a non-zero constant at the origin. The numerical solution is not reliable in this region. The `analytic' wave function is the variational estimate that minimizes the energy. 
The fit to data has a mild divergence at the origin.}

{\ }

	Our model does not predict the observed behavior of the parton  distributions 
for small $x_B$:
our probability distribution tends to a constant for small  $x_B$ (Sections~\ref{variational},~\ref{frobenius})
although due to numerical errors this is not evident in the numerical solution.
The observed distributions have an integrable singularity there: roughly
speaking, \m{\phi(x_B)\sim x_B^{-0.5}} for small \m{x_B}.The approximations we made clearly break down in the small \m{x_B}
region: travsverse momenta and sea quarks can no longer be ignored, indeed even gluons need to
be considered. We will study these effects in future publications. 

\pagebreak

{\ }

{\ }

{\ }

\section{Conclusion}
\label{conclusion}

	This thesis has been about the structure of the proton. The Deep Inelastic Scattering experiments showed that the proton is made of point-like constituents, called quarks or partons. The structure of the proton is described in terms of parton distribution functions. These are momentum-space probability distributions of quarks inside the proton. Parton distribution functions are the analogues of electron wave functions in an atom, and are of great importance in particle physics. They are measured in Deep Inelastic Scattering experiments. However, we do not yet have a theoretical understanding of the $x_B$ (momentum fraction or Bjorken Scaling variable) dependence of the parton distribution functions.

	In this thesis we solve the problem of predicting valence quark distributions in the proton. We have presented a model of interacting partons for the structure of baryons. The valence quarks interact through a linear potential in the null coordinate. This model can be derived from QCD in the approximation where transverse momenta are ignored. We obtain the valence quark wave function as the solution to a non-linear integral equation. We understand the behavior of the solution both analytically and numerically. Our prediction is compared with the baryon structure function extracted from global fits to Deep Inelastic Scattering data. The only parameter we can adjust is the fraction of baryon momentum carried by the valence partons. Our predictions agree well with data except for small values of the Bjorken Scaling variable, $x_B$, where our model is not expected to be accurate. We cannot ignore gluons, sea quarks and transverse momenta. We shall study these effects in future publications.

\pagebreak
\section*{Acknowledgements}
\label{acknowledgements}

	The author would like to thank his advisor Professor S. G. Rajeev for the wonderful opportunity to work on this problem and his insightful guidance. The author also thanks research associate Varghese John for useful discussions.

	This research was supported by the Department of Energy through 
the grant 
DE-FG02-91ER40685 and the Summer Reach program of the University of Rochester.


\end{document}